\documentclass[aps,preprint,unsortedaddress]{revtex4-1}

\usepackage{amsmath}%
\usepackage{graphicx}%
\usepackage{xcolor}%
\usepackage{dcolumn}%
\usepackage{bm}%
\usepackage{ulem}
\usepackage[uline]{hhtensor}
\usepackage{setspace}
\usepackage{siunitx}
\usepackage{lineno,hyperref}
\begin{document}
\title{The role of Si vacancies in the segregation of O, C and N at silicon grain boundaries: an ab initio study}
\author{Rita Maji}
\address{Dipartimento di Scienze e Metodi dell'Ingegneria, Universit{\`a} di Modena e Reggio Emilia,Via Amendola 2 Padiglione Tamburini , I-42122 Reggio Emilia, Italy}
\author{Julia Contreras-Garc\'{\i}a}
\address{Laboratoire de Chimie Th\'eorique, Sorbonne Universit\'e and CNRS  F-75005 Paris, France}
\author{Nathalie Capron} 
\address{Sorbonne Universit\'e, CNRS, Laboratoire de Chimie Physique Mati\`ere et Rayonnement, UMR 7614, F-75005 Paris, France.}
\author{Elena Degoli}
\email{elena.degoli@unimore.it}
\address{Dipartimento di Scienze e Metodi dell'Ingegneria, Universit{\`a} di Modena e Reggio Emilia, Via Amendola 2 Padiglione Morselli, I-42122 Reggio Emilia, Italy, \\ 
Centro Interdipartimentale En$\&$Tech, Via Amendola 2 Padiglione Morselli, I-42122 Reggio Emilia, Italy, \\
Centro S3, Istituto Nanoscienze-Consiglio Nazionale delle Ricerche (CNR-NANO),Via Campi 213/A, 41125 Modena, Italy}
\author{Eleonora Luppi}
\email{eleonora.luppi@sorbonne-universite.fr}
\address{Laboratoire de Chimie Th\'eorique, Sorbonne Universit\'e and CNRS  F-75005 Paris, France}

\date{\today}


\begin{abstract}
Grain boundaries (GBs) are defects originating in multi-crystalline silicon during crystal growth for device Si solar cell fabrication. The presence of GBs changes the coordination of Si making advantageous for charge carriers to recombine which brings a significant reduction of carrier lifetimes. Therefore, GBs can be highly detrimental for the device performances. Furthermore, GBs easily form vacancies with deep defect electronic states and are also preferential segregation sites for various impurity species as C, N and O. We studied from first-principles the correlation between structural, energetics and electronic properties of the $\Sigma$3\{111\} Si GB, with and without vacancies, and the segregation of C, N and O atoms. C and O atoms strongly increase their ability to segregate when vacancies are present. However, the electronic properties of the $\Sigma$3\{111\} Si GB are not affected by the presence of O while they can strongly change in the case of C. For N atoms it is not possible to find a clear trend in the energetics and electronic properties both with and without vacancies in the GB. In fact, as N is not isovalent with Si, as C and O, it is more flexible in finding new chemical arrangements in the GB structure.  This implies a stronger difficulty in controlling the properties of the material in the presence of N impurity atoms compared to C and O impurities.
\end{abstract}



\maketitle

\section{Introduction}
\label{sec:introduction}

Grain boundaries (GBs) are defects originating in multi-crystalline silicon during crystal growth for device fabrication of Si solar cells. \cite{Sisolarcell2009review,Sisolarcell2008review,Sisolarcellthinfilm2004,Sisolarcellbook2019} 
GBs are highly detrimental for device performances  \cite{GBeffect1997,GBeffect1990,GBeffect2012} as Si coordination is changed. This makes advantageous charge carriers recombination which induces a significant reduction of carrier lifetimes. \cite{lifetime2005,PhysRevLett.115.235502,Feng2009,1.4932203} 

Furthermore, GBs easily form vacancies with deep defect electronic states \cite{Feng2009} and are also preferential segregation sites for various impurity species.\cite{ZHAO2017599,PhysRevB.91.035309,ohnoapl2013,KasinnoJAP13} In fact, during the crystal growth process atoms such as C, O and N can segregate at GBs and play an important role in the formation of precipitates such as SiO$_2$, SiC and Si$_3$N$_4$. Therefore C, O and N atomic species can alter the electronic and the mechanical properties of Si solar cells. \cite{KasinnoJAP13,YuAndreyJAP2015,OhnoAPL2017,seagerAnnReVMatSci85,OhnoAPL15,ShiJAP2010,PhysRevLett.121.015702} Recombination and segregation activity can have a substantial detrimental impact on the conversion efficiency of the solar cells. \cite{YuAndreyJAP2015,Peaker2012} 

O atoms are inevitably introduced during solar-cell crystal growth which usually form precipitate recombination centers with different morphology. \cite{Chen2011,OhnoAPL15,1.1578699,1.98331}  Complex mechanisms control the segregation of oxygen atoms at Si GBs which is influenced by the size and the orientation of the grains and also by the presence of vacancies and strain in the GBs.\cite{KasinnoJAP13,YuAndreyJAP2015,OhnoAPL2017,seagerAnnReVMatSci85,OhnoAPL15,ShiJAP2010,PhysRevLett.121.015702,1.1578699,1.98331}
Segregation of C atoms at GBs can originate amorphous SiC clusters and filaments which impact the quality of the multi-crystalline silicon. \cite{Chen2015,pssr.201600354,KasinnoJAP13,1.98331} Moreover, the presence of C atoms strongly enhances the precipitation of O during thermal annealing. \cite{ZHAO2017599} Also N atoms segregate at Si GBs reducing the size of the voids and modifying the mechanical properties of the material. \cite{PhysRevB.62.1851,1.126760} 

In this work, we studied from first-principles the correlation between structural and electronic properties of the $\Sigma$3\{111\} Si GB, with and without vacancies, and the segregation of C, N and O atoms.
We discussed different factors that can cause these atoms to segregate at GB and which could then prevent from engineering the GB for optimised device performances. The electronic properties have also been investigated by characterising the origin of the new energy levels due to the presence of GBs, vacancies and/or the impurity atoms. Understanding how to characterise the structural and the electronic properties of a GB can be fundamental to understand and control the properties of a device.

This paper is organized as follows: In Sec.\ref{sec:method}, we briefly discuss
the methodology used to perform the calculations; in Sec.\ref{sec:methogb}, we
describe the GB structures studied; in Sec.\ref{sec:results}, we report the results
and we discuss the role of C, O, and N impurities in  $\Sigma$3\{111\} Si GB
without and with vacancies; and finally, in Sec.\ref{sec:conclude}, we conclude our work.

\section{Methodology}\label{sec:method}

The calculations were performed using density functional theory (DFT) as implemented in the plane-wave based Vienna Ab initio Simulation Package (VASP).\cite{Hafner, Kresse} We employed the generalised gradient approximation PBE for the exchange-correlation functional \cite{PhysRevLett.77.3865} and projector augmented-wave (PAW) pseudopotentials with a cutoff of 400 eV. K-points sampling within the Monkhorst Pack scheme \cite{Monkhorst} was used for integration of Brillouin-zone together with the linear tetrahedron method including Bl\"ochl corrections. \cite{PhysRevB.49.16223} In particular, we used a k-mesh of 3$\times$3$\times$3 to calculate energetics of the structures and a k-mesh of 7$\times$7$\times$7 to calculate their density of states (DOS). For the structural optimisation, we used a force threshold value of 10$^{-2}$ eV/\AA{} per atom. For all the systems the total energy is calculated as the total energy of the unit cell where all the atoms are interacting. Moreover, the pressure of the systems is calculated directly from VASP and it is the hydrostatic pressure of a cell of a given volume. Finally, in these calculations we did not include the spin polarization \cite{Zhu2015,PhysRevB.80.144112} which will be studied in a future work.

\section{Grain boundary structures}\label{sec:methogb}

\begin{figure*}[h!]
\centering
\includegraphics[scale=0.45]{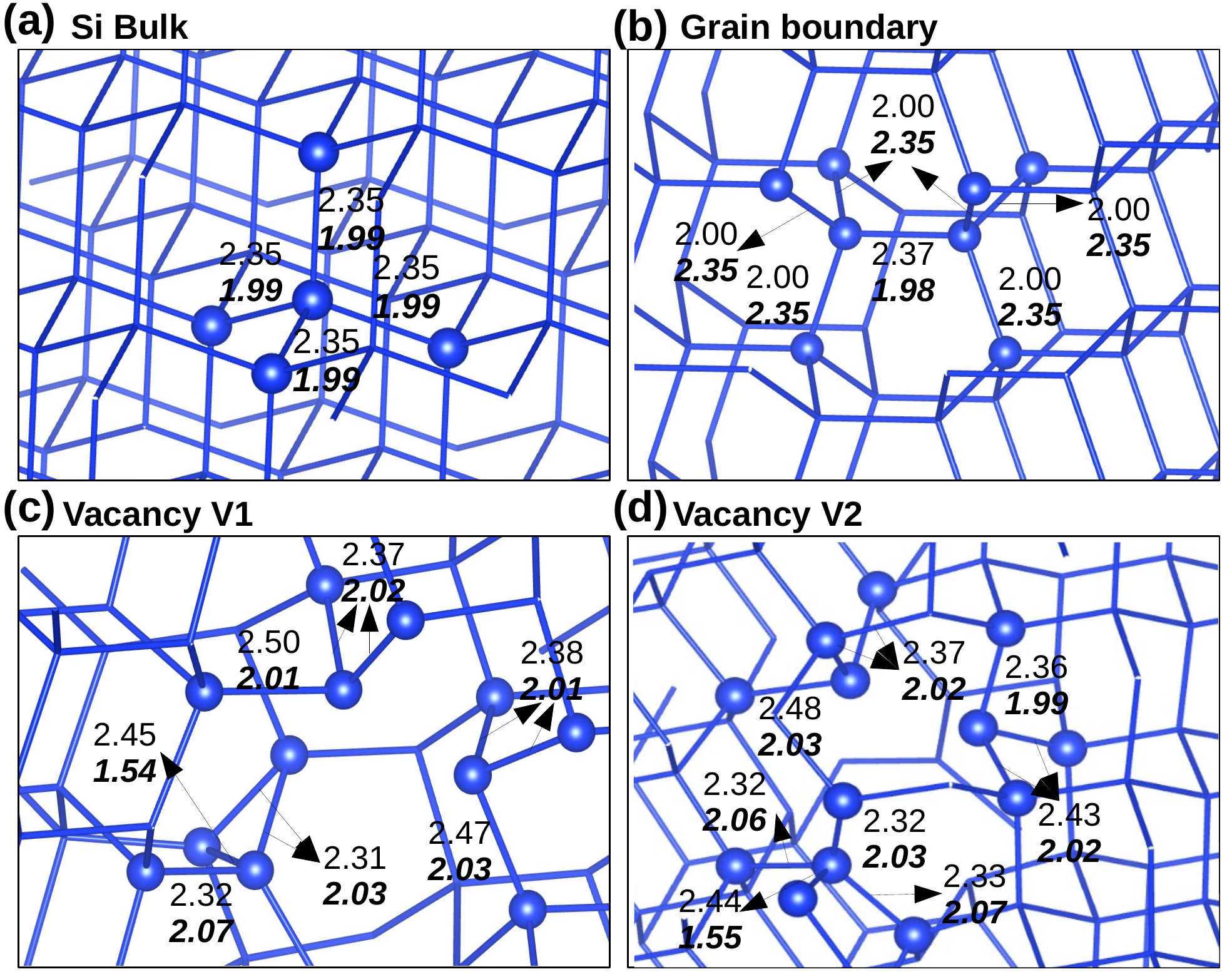}
\caption{Bond lengths (regular) and bond charges (bold italic)  in Si bulk (a), in $\Sigma$3\{111\} Si GB (b), in $\Sigma$3\{111\} Si GB with vacancy V1 (c) and in $\Sigma$3\{111\} Si GB with vacancy V2 (d).}
\label{bulkGBV1V2}
\end{figure*}

The $\Sigma$3\{111\} Si GB consists of two Si grains, misoriented by an angle $\Omega = 60^{\circ}$ which form an interface along the crystallographic plane \{111\} (coincidence site lattice). To describe the grain, we used an orthorhombic supercell ($a$ $\ne$ $b$ $\ne$ $c$ and $\alpha$ = $\beta$ = $\gamma$= $90^{\circ}$) composed of 96 Si atoms generated with {\it GB Studio program}. \cite{GBStudio2006}. The lattice parameters are $a$=13.30~\AA, $b$=7.68~\AA~and $c$=18.81~\AA.  \cite{MAJI2021116477} The formation energy of the $\Sigma$3\{111\} Si GB is $E^{\text{f}}_{\text{GB}} = 0.002$ eV/\AA$^{2}$ ($E^{\text{f}}_{\text{GB}} = 0.05$ J/m$^{2}$) \cite{MAJI2021116477} which indicates that this GB has a very regular structure, i.e. bond lengths and angles are close to the Si bulk. \cite{ZHAO2017599,PhysRevB.91.035309} This is also the most stable GB since there are no dangling bonds and little bond distortion. \cite{SARAU20112264} The representation of the $\Sigma$3\{111\} GB is in panel (b) of Fig.~\ref{bulkGBV1V2} while panel (a) of Fig.~\ref{bulkGBV1V2} shows the structure of Si bulk that we took as the reference to compare our results. In this case, we used a cubic supercell ($a=b=c$ and $\alpha$ = $\beta$ = $\gamma$= $90^{\circ}$) of 64 atoms with $a$=10.86~\AA. This value for $a$ was deduced from the experimental lattice constant 5.431 \AA\,\,for a face-centered cubic unit cell of two Si atoms. \cite{PhysRevB.32.3792} We also report the calculated Si bulk modulus $B$ = 95.4 GPa and the elastic tensor components $c_{11}$ = 159.8 GPa  and $c_{12}$ = 63.2 GPa. These values are in good agreement with other theoretical calculations and with the experimental values $B$ = 99.2 GPa, $c_{11}$ = 167.5 GPa  and $c_{12}$ = 65.0 GPa. \cite{PhysRevB.32.3792}

\begin{figure*}[h!]
\centering
\includegraphics[scale=0.27]{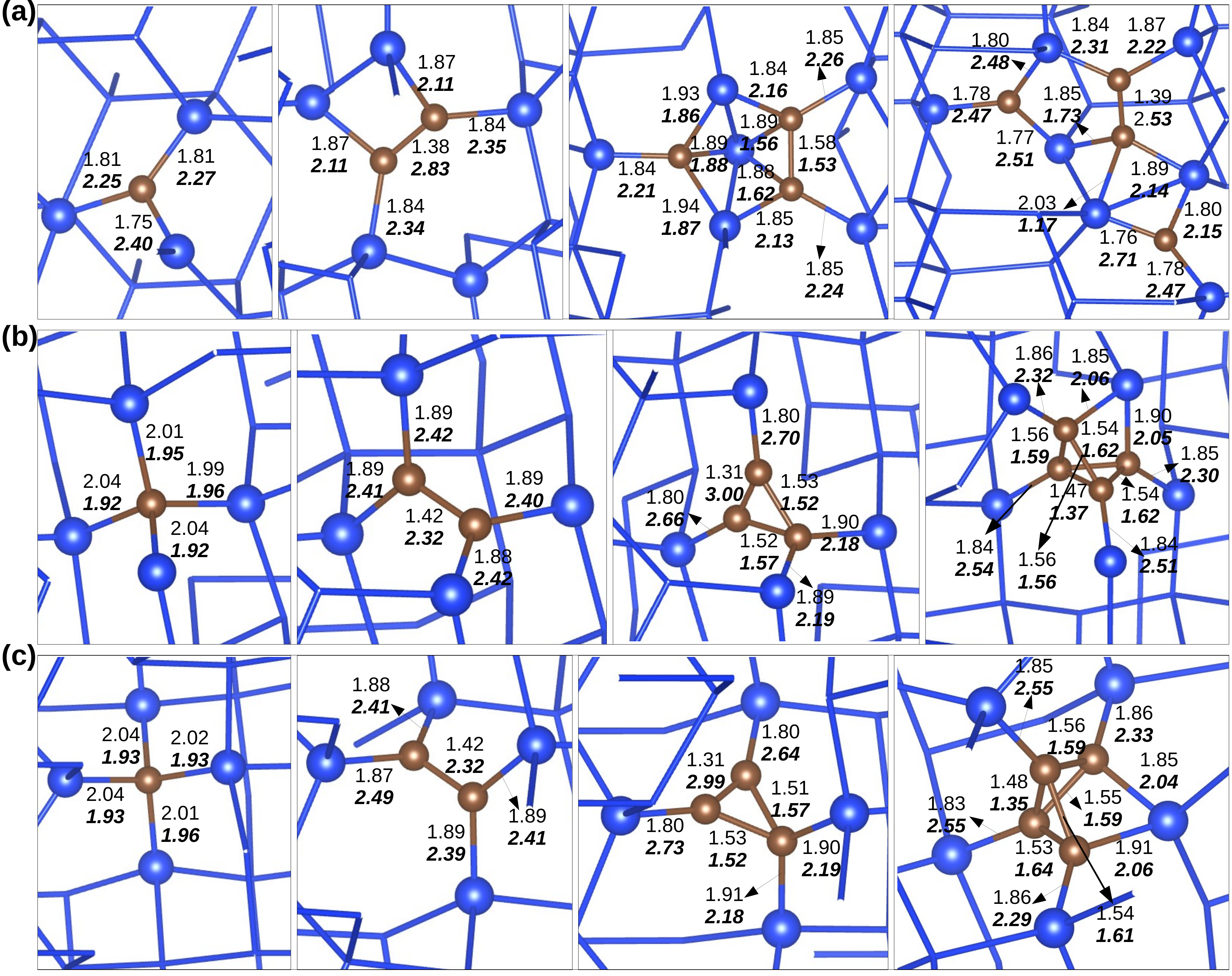}
\caption{Bond lengths (regular) and bond charges (bold italic) for interstitial C atoms (brown balls) in $\Sigma$3\{111\} Si GB (a), in $\Sigma$3\{111\} Si GB with vacancy V1 (b) and in $\Sigma$3\{111\} Si GB with vacancy V2 (c). From left to right the number of C atoms in the cell increases from 1 to 4.}
\label{gb_C}
\end{figure*}

\begin{figure*}[h!]
\centering
\includegraphics[scale=0.25]{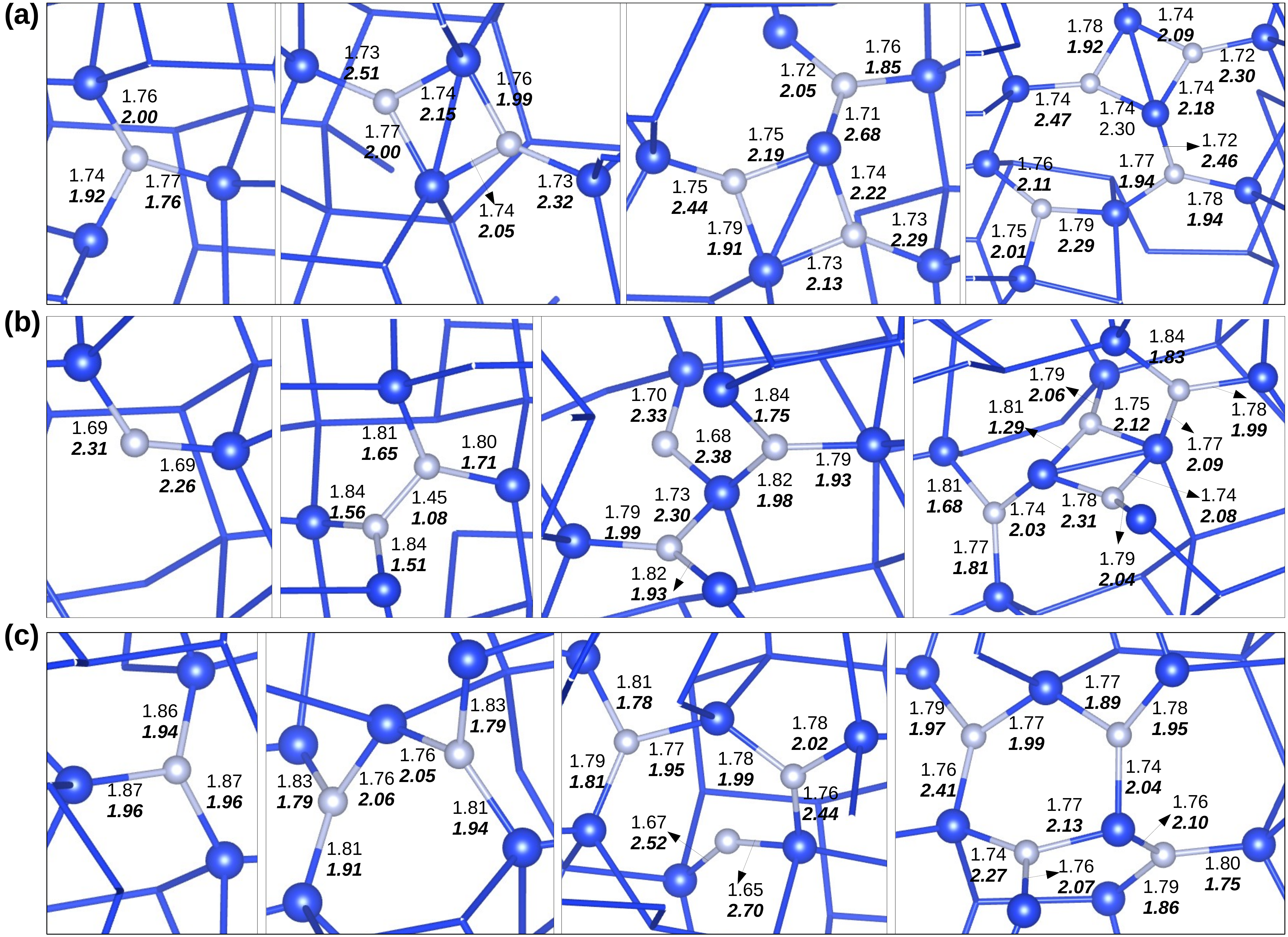}
\caption{Bond lengths (regular) and bond charges (bold italic) for interstitial N atoms (grey balls) in $\Sigma$3\{111\} Si GB (a), in $\Sigma$3\{111\} Si GB with vacancy V1 (b) and in $\Sigma$3\{111\} Si GB with vacancy V2 (c). From left to right the number of N atoms in the cell increases from 1 to 4.}
\label{gb_N}
\end{figure*}

\begin{figure*}[h!]
\centering
\includegraphics[scale=0.25]{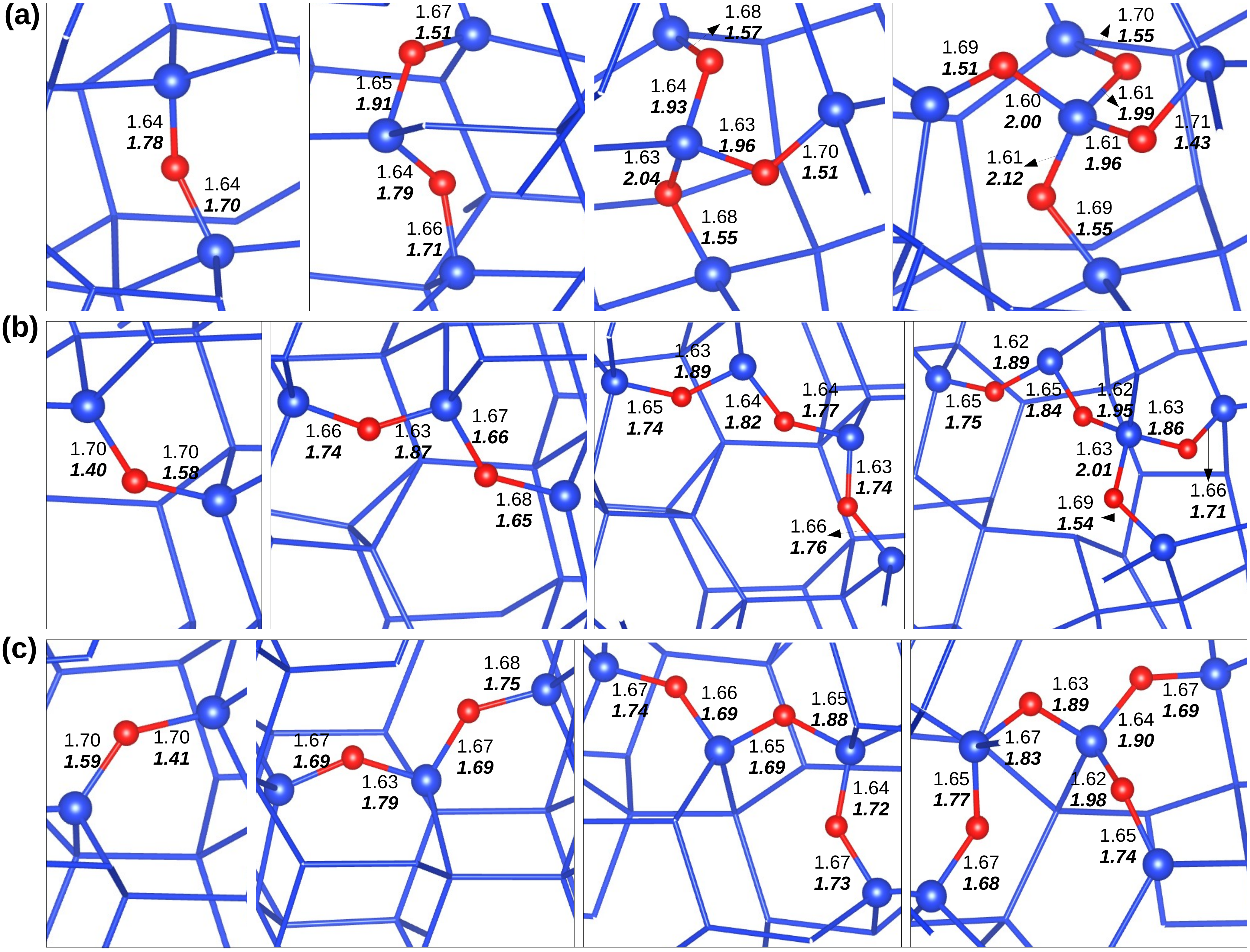}
\caption{Bond lengths (regular) and bond charges (bold italic) for interstitial O (red balls) atoms in $\Sigma$3\{111\} Si GB (a), in $\Sigma$3\{111\} Si GB with vacancy V1 (b) and in $\Sigma$3\{111\} Si GB with vacancy V2 (c). From left to right the number of O atoms in the cell increases from 1 to 4.}
\label{gb_O}
\end{figure*}

Then, we created a vacancy in the $\Sigma$3\{111\} Si GB. There exists only two distinct positions of Si where vacancies can be created. \cite{MAJI2021116477} These vacancies are indicated throughout the paper as V1 and V2 and are shown respectively in panel (c) and (d) of Fig.~\ref{bulkGBV1V2}. The formation energy of the vacancy V1 in the GB is $E^{\text{f}}_{\text{VGB}}$ = 2.91 eV while for the vacancy V2 we found $E^{\text{f}}_{\text{VGB}} = 3.04$ eV. \cite{MAJI2021116477} The presence of the vacancy induces coordination defects, and in the case of V1 and V2 we have two Si atoms that are threefold coordinated while all the other Si atoms are fourfold coordinated.  
\cite{Feng2009,PhysRevB.37.7482,PhysRevB.58.1318,ZHANG2020144437,YuAndreyJAP2015,MAJI2021116477} 

To study the role of C, N and O segregation we then inserted these atoms as interstitials in the Si GB with and without vacancies. Different configurations were found and in Figs.~\ref{gb_C},\ref{gb_N} and \ref{gb_O} are reported only those that corresponds to the lowest total energy with a number of impurities increasing from 1 to 4.

To study the structural changes induced by the impurities in the GB we analysed the correlation between the bond lengths and the bond charges. \cite{SJ92} In order to dissect bond charges we have resorted to the Electron Localization Function (ELF). \cite{KS96,GS02} ELF permits to know when a bond is present in the system, as it allows to divide the system into three dimensional regions ($\Omega$) that have a clear chemical meaning (cores, bonds, lone pairs, etc). Hence, integrating the electron density, over the bonding region $\Omega$, enables to retrieve the bond charges which is defined as $q=\int_\Omega n({\bf r}) \text{d}{\bf r}$.

Starting from bulk silicon in Fig.~\ref{bulkGBV1V2} we observe a regular tetrahedral coordination with a Si-Si covalent bond holding 2 electrons (2e). When the grain boundary is formed the bonding is very much preserved and we observe the appearance of an elongated Si-Si bond with a slightly smaller charge.
The introduction of vacancies leads much more important bonding changes. Two 3-fold coordinated Si atoms appear in V1 and V2 configurations leading to an overall increase of the charge in the neighbouring Si-Si bonds (2.01e-2.07e). Moreover, both in V1 and in V2 an unstable triangular coordination with a small charge of 1.54e and 1.55e respectively appears.

In panel (a) of Fig.~\ref{gb_C} interstitial C atoms in $\Sigma$3\{111\} Si GB are shown. C is isovalent with Si and tends to form simple bonds with Si. The bond lengths are of the order of the bulk SiC which is 1.88 \AA.
In the case of one C atom, the C is three-fold coordinated. Instead, the presence of two C atoms creates a double bond between the C atoms and no unpaired electron(s) are found in the structure. 
When a greater number of atoms are introduced, they cannot be accommodated all together anymore and need to occupy voids with at most one C-C bond. This leads to unstable conformations: five-fold Si coordination for 3C and a 3-fold Si coordination for 4C atoms.  

When the voids are introduced through vacancies (panels (b,c) of Fig.~\ref{gb_C} ), higher number of carbon atoms can be accommodated more easily, leading to well known carbon arrangements even for high C contents: ciclopropene for 3C and a bicycle for 4C. Nonetheless, these two molecule configurations are known to be highly strained and C-C bonds with low occupations (e.g. 1.3-1.5) are found. Strained bonds are also confirmed by the position of the bonding charges outside the bonding line. Between all the structures in V1 and V2 those with two C atoms are particularly unstable. In fact, the C-C bond length is shorter than a simple bond but longer than a double bond which makes the electrons in these structures more delocalised, and as we will see, the system is gapless.

In panel (a) of Fig.~\ref{gb_N} interstitial N atoms in $\Sigma$3\{111\} Si GB are shown, while in panels (b,c) of the same figure  interstitial N atoms are shown in the presence of vacancies V1 and V2. When inserting N in the GB where all the Si are 4 coordinated the N atom always try to get the natural coordination of N being 3. However, in the case of 1N and 3N atoms we found a Si that is 3 coordinated too which makes these structures, as we will see, gapless. In the case of 2N and 4N atoms a flat N 3-coordination squared structure (diborane type structure) is formed where Si can be in some cases over-coordinated. In the case of vacancies we have the same behaviour except that in some cases N is 2-fold coordinated and in the case of V2 we get an hexagonal structure.

In panel (a) of Fig.~\ref{gb_O} interstitial O atoms in $\Sigma$3\{111\} Si GB are shown, while in panel (b,c) of the same figure  interstitial O atoms are shown in the presence of vacancies V1 and V2. When inserting O in the GB where all the Si are 4 coordinated the O atoms always segregate at  bond-centered position between two Si atoms, restoring the configuration of SiO$_2$. Therefore, the optimised structures have no coordination defects as all Si atoms restored their tetrahedral coordination. \cite{OhnoAPL2017}  When vacancies are present the behaviour of O atoms continues to be two fold Si-O coordinated, which favours the formation of Si-O chains that enable to “repair” the defaults. In general, smaller Si-O bond charges are found due to the stretched bond to fit the void. \cite{MAJI2021116477}

\begin{figure}[t!]
\centering
\includegraphics[scale=0.5]{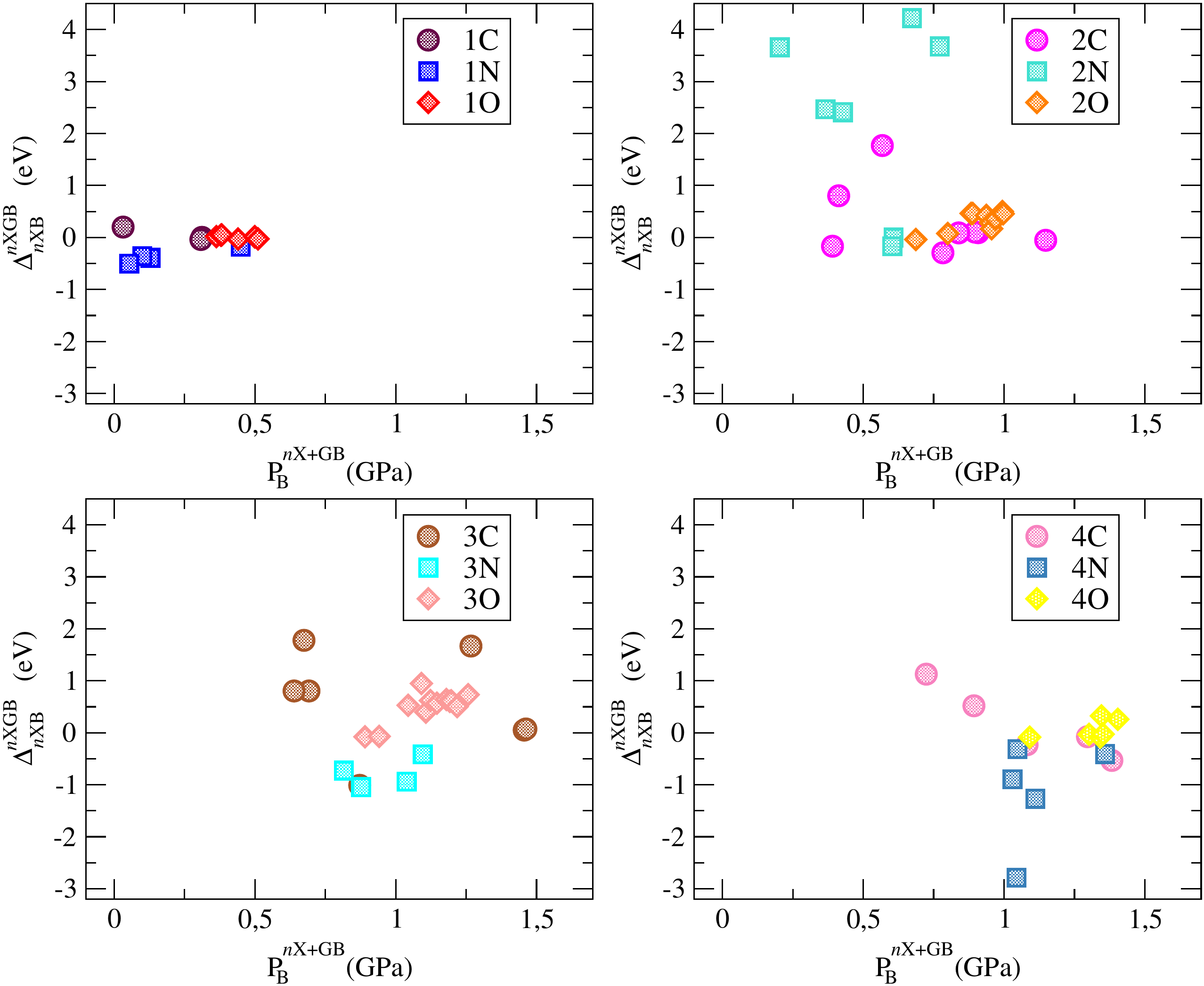}
\caption{Segregation energy $\Delta^{n\text{XGB}}_{n\text{XB}}$ (eV) as a function of P$^{n\text{X+GB}}_\text{B}$ (GPa). The label X stands for the type of atoms C, N and O, while $n$ indicates the number of atomic impurities. The atoms C, N and O are interstitials in the GB.}
\label{seg_cno_int}
\end{figure}

\section{Results and discussion}\label{sec:results}

\subsection{Carbon, oxygen and nitrogen atoms in $\Sigma$3\{111\} Si GB}
\label{XXX}

To investigate the interaction between the $\Sigma$3\{111\} Si GB and the C, N and O atoms we calculated the segregation energy as 

\begin{equation}
\Delta^{n\text{XGB}}_{n\text{XB}} = E^{n\text{XGB}} - E^{n\text{XB}},
\label{bulkseg} 
\end{equation}
where the label X stands for the C, N or O atoms, $E^{n\text{XGB}}$ and $E^{n\text{XB}}$ are the impurity energies of the X atom in the GB and in the Si bulk, respectively, which are calculated as

\begin{equation}
E^{n\text{XGB}} = E_{n\text{X+GB}} - E_{\text{GB}}  - n \mu_{\text{X}},
\label{EnXGB} 
\end{equation}

\begin{equation}
E^{n\text{XB}} =  E_{n\text{X+B}} - E_{\text{B}} - n \mu_{\text{X}}.
\label{EnXB} 
\end{equation}

$E_{n\text{X+GB}}$ is the total energy of the GB containing $n$ number of X atoms, $E_{\text{GB}}$ is the total energy of the GB,  $\mu_{\text{X}}$ is the chemical potential of the X atom, $E_{n\text{X+B}}$ is the total energy of the Si bulk containing $n$ number of X atoms \cite{PhysRevB.80.144112} and $E_{\text{B}}$ is the total energy of Si bulk. 

In Fig.~(\ref{seg_cno_int}) we show the segregation energy $\Delta^{n\text{XGB}}_{n\text{XB}}$ (eV) for C, N and O atoms as a function of the pressure $P^{n\text{X+GB}}_{\text{B}}=P_{n\text{X+GB}} - P_{\text{B}}$ (GPa), calculated as the difference between the pressure of the GB containing a number $n$ of X atoms ($P_{n\text{X+GB}}$) and the pressure of the Si bulk ($P_{\text{B}}=1.973$ GPa).

\begin{table*}[b!]
\begin{center}
\begin{tabular}{ |c |c |c |c |c |c|c|c|c| } 
\hline
X &$P^{\text{1X+GB}}_{\text{B}}$ & $\Delta^{\text{1XGB}}_{\text{1XB}}$ &$P^{\text{2X+GB}}_{\text{B}}$ & $\Delta^{\text{2XGB}}_{\text{2XB}}$& $P^{\text{3X+GB}}_{\text{B}}$ &  $\Delta^{\text{3XGB}}_{\text{3XB}}$&$P^{\text{4X+GB}}_{\text{B}}$ &  $\Delta^{\text{4XGB}}_{\text{4XB}}$ \\ 
\hline
\hline 
C  & 0.308  & -0.039 &  0.783 & -0.295  &   0.872  &-1.012& 1.383 &-0.532  \\ 
N  & 0.054 & -0.503 &  0.604 & -0.169 &  0.877  & -1.046  &1.044  &-2.786  \\ 
O  & 0.440 & -0.030 &  0.687 & -0.039  &0.891 & -0.078  & 1.091& -0.085 \\ 
\hline
\end{tabular}
\end{center}
\caption{Segregation energy $\Delta^{n\text{XGB}}_{n\text{XB}}$ (eV) as a function of $P^{n\text{X+GB}}_{\text{B}}$ (GPa). The label X stands for the type of atoms C, N and O while $n$ indicate the number atomic impurities. The values are reported for the LE (lowest total energy) structures.}
\label{seg_cno_int_le}
\end{table*}

We obtained that the segregation energy of O is of the order of $10^{-2}$ eV, which is quite small. This confirms that $\Sigma$3\{111\} Si GB has no gettering ability to solute oxygen atoms. \cite{ohnoapl2013,OhnoAPL2016,SARAU20112264} The segregation energy of C is slightly larger than that of O and becomes more favourable when the number of C atoms increases. The case of N is different. Depending on the number of atoms introduced in the GB we obtain a segregation energy that can oscillate between positive and negative values. In particular, when it is favourable (i.e. negative values) the segregation of N can be much stronger than the one for C and O. In Tab.~(\ref{seg_cno_int_le}) we report the segregation energy $\Delta^{n\text{XGB}}_{n\text{XB}}$ (eV) as a function of $P^{n\text{X+GB}}_{\text{B}}$ (GPa) for C, N and O atoms for those structures that have the lowest total energy ($E_{n\text{X+GB}}$). 

 
It is then important to analyse the electronic properties of these systems. In fact, despite the segregation of these impurities at the GB the change in the electronic properties is not obvious. To make this study we continued to use the PBE functional. Concerning the PBE, it is well known that this functional underestimates the electronic band gaps. For Si bulk we obtain 0.574 eV (see Tab.~(\ref{Eg})) compared with the experimental value of 1.25 eV. \cite{lupp+08prb}. Hybrids functionals or GW corrections can be used to obtain higher accuracy, but the computational cost is very high for the systems studied here. However, despite the underestimation of the absolute values of the gaps, the trend of the electronic properties is also valid in PBE.\cite{PhysRevB.92.081204,Zhu2015}

\begin{table*}[h!]
\begin{center}
\begin{tabular}{ |c |c |c |c | c | } 
\hline
&$E^{\text{Bulk}}_{\text{g}}$ & $E^{\text{GB}}_{\text{g}}$ &$E^{\text{V$_1$GB}}_{\text{g}}$  & $E^{\text{V$_2$GB}}_{\text{g}}$  \\ 
\hline
\hline 
Si & 0.5736  & 0.5694 &  0.4359 & 0.4364 \\ 
\hline 
\hline
X &$E^{\text{1X+GB}}_{\text{g}}$ & $E^{\text{2X+GB}}_{\text{g}}$ & $E^{\text{3X+GB}}_{\text{g}}$ & $E^{\text{4X+GB}}_{\text{g}}$ \\ 
\hline
\hline 
C & 0.0477  & 0.5235 &  0.4729 & 0.1319  \\ 
N & - & 0.5444 &  - & 0.6274    \\ 
O & 0.5506 & 0.6073 &  0.6185 & 0.5569 \\ 
\hline
\hline
X&$E^{\text{1X+V1GB}}_{\text{g}}$ & $E^{\text{2X+V1GB}}_{\text{g}}$ & $E^{\text{3X+V1GB}}_{\text{g}}$ & $E^{\text{4X+V1GB}}_{\text{g}}$ \\ 
\hline
\hline 
C & 0.3587  & - &  0.6066 & 0.6502    \\ 
N & - & 0.2439 & - & 0.3966   \\ 
O & 0.6274 & 0.6353 &  0.6306 &0.6850 \\ 
\hline
\hline
X&$E^{\text{1X+V2GB}}_{\text{g}}$ &  $E^{\text{2X+V2GB}}_{\text{g}}$  &$E^{\text{3X+V2GB}}_{\text{g}}$ & $E^{\text{4X+V2GB}}_{\text{g}}$ \\ 
\hline
\hline 
C & 0.4014  & - &  0.5578& 0.5978   \\ 
N & - & 0.6191 &  - & 0.6397    \\ 
O & 0.6283 & 0.6292 &  0.6319 & 0.6989 \\ 
\hline
\end{tabular}
\end{center}
\caption{Energy gaps (eV) of Si bulk, $\Sigma$3\{111\} Si GB, $\Sigma$3\{111\} Si GB with C, N and O interstitial atoms, $\Sigma$3\{111\} Si GB with vacancies V1 and V2 and $\Sigma$3\{111\} Si GB with vacancies V1 and V2 with C, N and O interstitial atoms. The values are reported for the LE structures.}
\label{Eg}
\end{table*}

\begin{figure}[b!]
\centering
\includegraphics[scale=0.3]{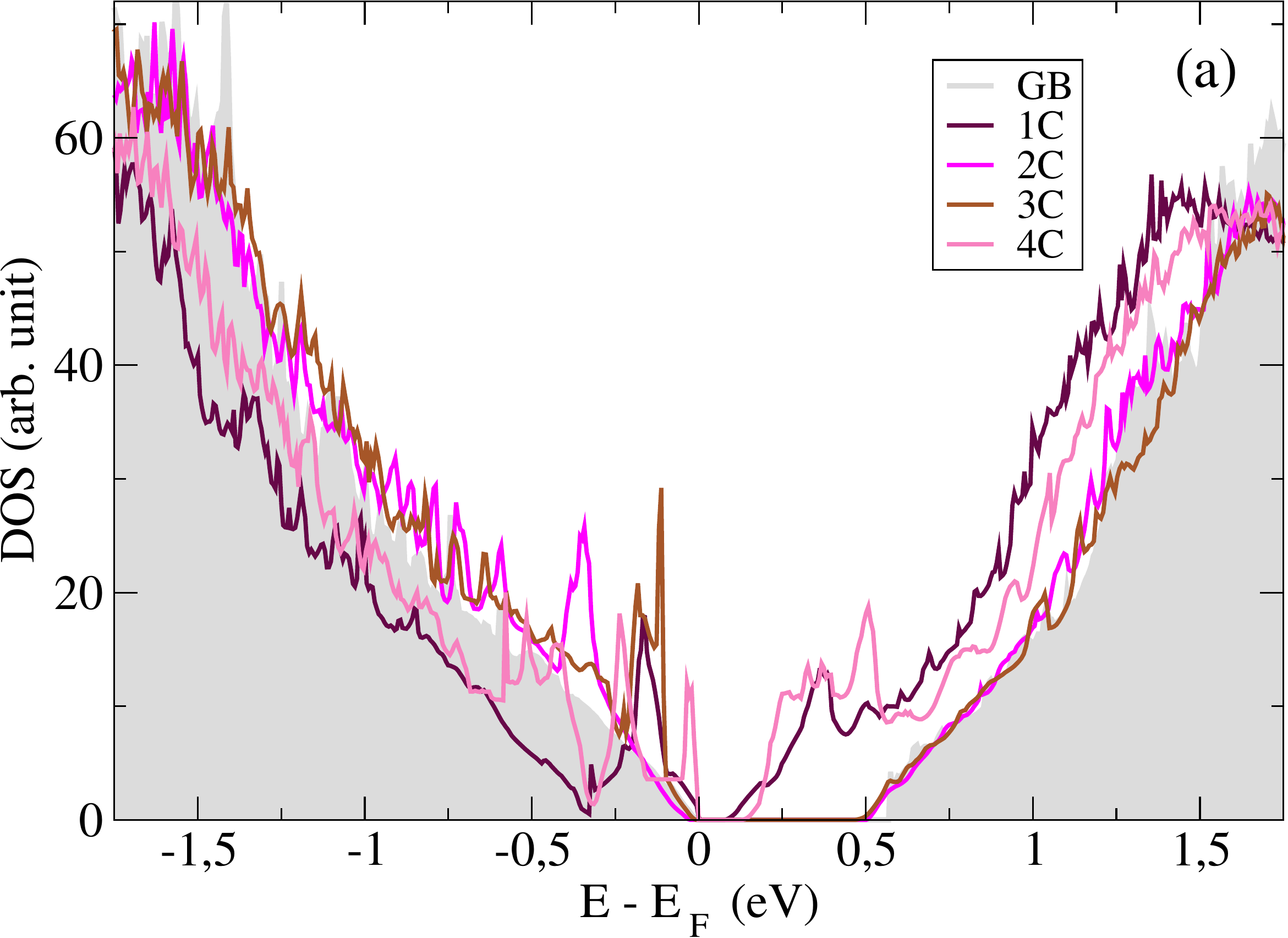}
\includegraphics[scale=0.3]{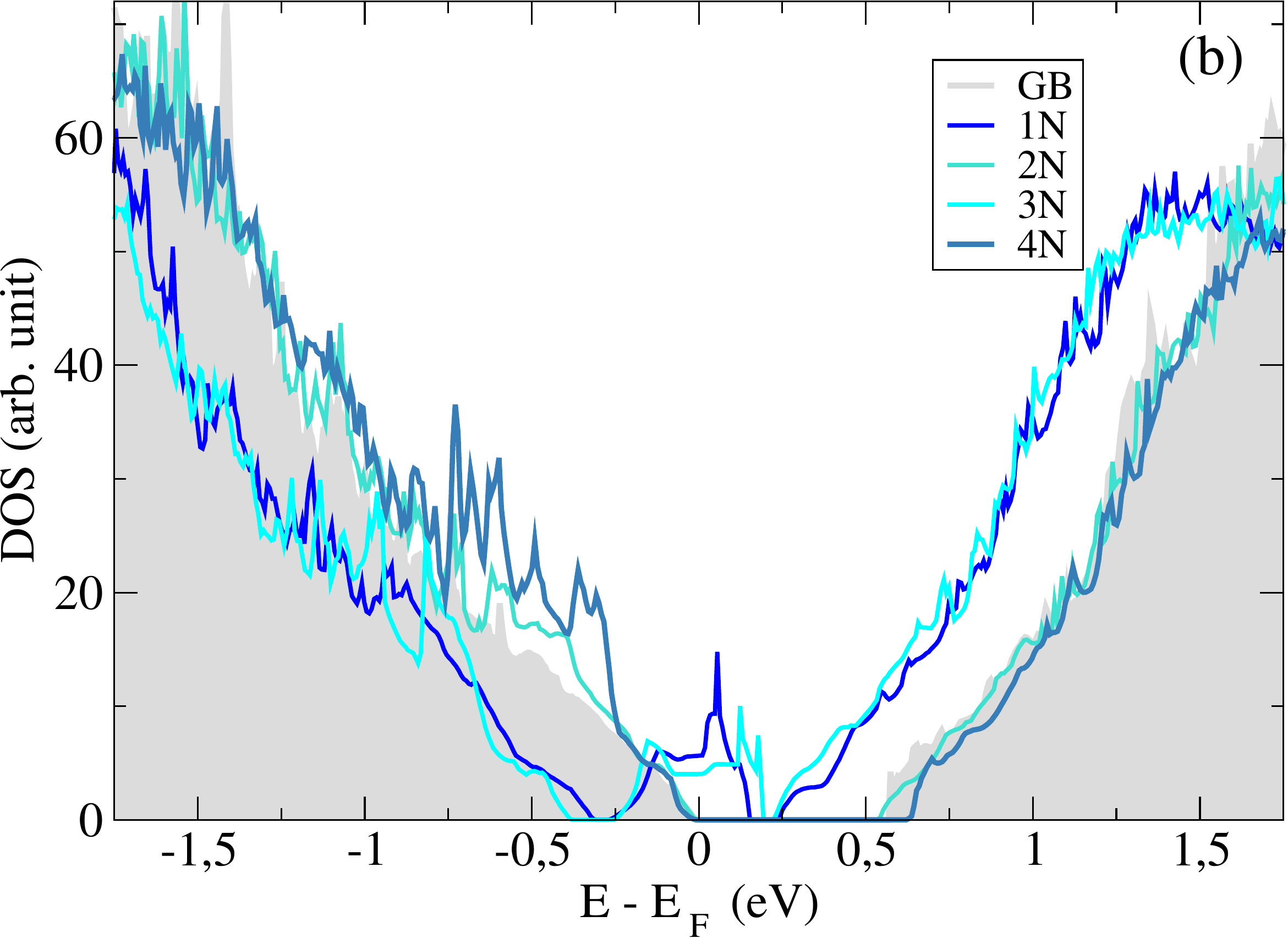}
\includegraphics[scale=0.3]{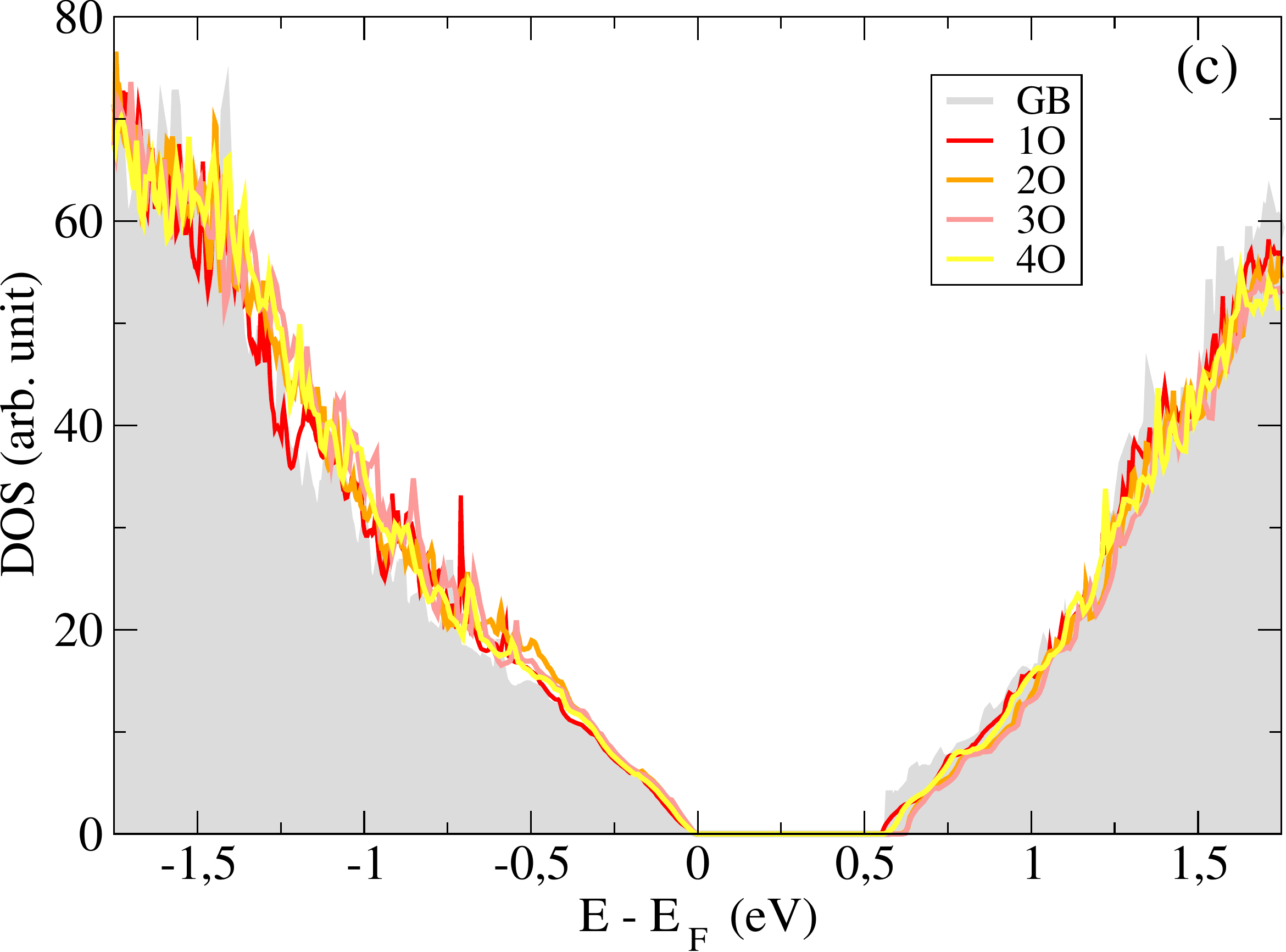}
\caption{Density of states (DOS) of $\Sigma$3\{111\} Si GB with and without C (a), N (b) and O (c) interstitial atoms.}
\label{dos_int}
\end{figure}

In  panel (a) of Fig.~(\ref{dos_int}) we show the density of states (DOS) for the $\Sigma$3\{111\} Si GB with and without interstitial C, N and O atoms. The lowest total energy structures are plotted. In the case of 1C and 4C the gap becomes much smaller than in pure GB (see Tab.~(\ref{Eg})). In both cases we have a C that is 3-fold coordinated with a Si that is 3-fold coordinated too. This configuration delocalise the electrons diminishing the gap and creates new states at the top of the valence bond due to C and at the bottom of the conduction band due to Si. In the case of 2C the C atoms are still 3-fold coordinated but they form a double bond (1.38\AA) as clearly shown from the bond charge of 2.83e, while in the case of 3C both C and Si atoms are always 4-fold coordinated restoring the SiC structure. The structures of 2C and 3C give a gap and a DOS similar to those obtained for the pure GB.

In panel (b) of Fig.~(\ref{dos_int}) we show the DOS for interstitial N atoms. The presence of N tends to break Si-Si bonds in order to be three coordinated and the electron pairs are localised on the nitrogen. . The analysis of the 1N and 3N structures in the Sec.\ref{sec:methogb} showed that there is a Si which is 3-fold coordinated. The unpaired electron from Si is delocalised making the 1N and the 3N structures gapless. Instead, the electronic properties are almost unchanged with respect to the GB without impurities in the case of 2N and 4N. This electronic stability reflects the correlation between structure and covalent bonding (Sec.\ref{sec:methogb}). This behaviour is even more clearly observed in the case of O atoms.

\subsection{Carbon, oxygen and nitrogen atoms in $\Sigma$3\{111\} Si GB with vacancies}

The interaction of C, O and N impurities in the $\Sigma$3\{111\} Si GB with a vacancy was quantified by calculating the segregation energy of the impurity atoms  as 

\begin{equation}
\Delta^{n\text{XVGB}}_{n\text{XGB}} = E^{n\text{XVGB}} - E^{n\text{XGB}}, 
\label{deltanXVGB} 
\end{equation}
where $E^{n\text{XVGB}}$ and $E^{n\text{XGB}}$ are the impurity energies respectively in the GB with and without the vacancy. The energy  $E^{n\text{XVGB}}$ is calculated as
\begin{equation}
E^{n\text{XVGB}} = E_{n\text{X+VGB}}- E_{\text{VGB}}  - n \mu_{\text{X}}
\label{EnXVGB} 
\end{equation}
where $E_{n\text{X+VGB}}$ is the energy of the GB with the vacancy with a number $n$ of atoms of the type X, $E_{\text{VGB}}$ is the energy of the GB with a vacancy. The energy $E^{n\text{XGB}}$ is calculated as
\begin{equation}
E^{n\text{XGB}} = E_{n\text{X+GB}} - E_{\text{GB}}  - n \mu_{\text{X}},
\label{EnXGB} 
\end{equation}
where $E_{n\text{X+GB}}$ is the energy of the GB containing $n$ atoms of the type X, $E_{\text{GB}}$ is the energy of the GB and $\mu_{\text{X}}$ is  the chemical potential of the impurity atom X.

\begin{figure}[t!]
\centering
\includegraphics[scale=0.5]{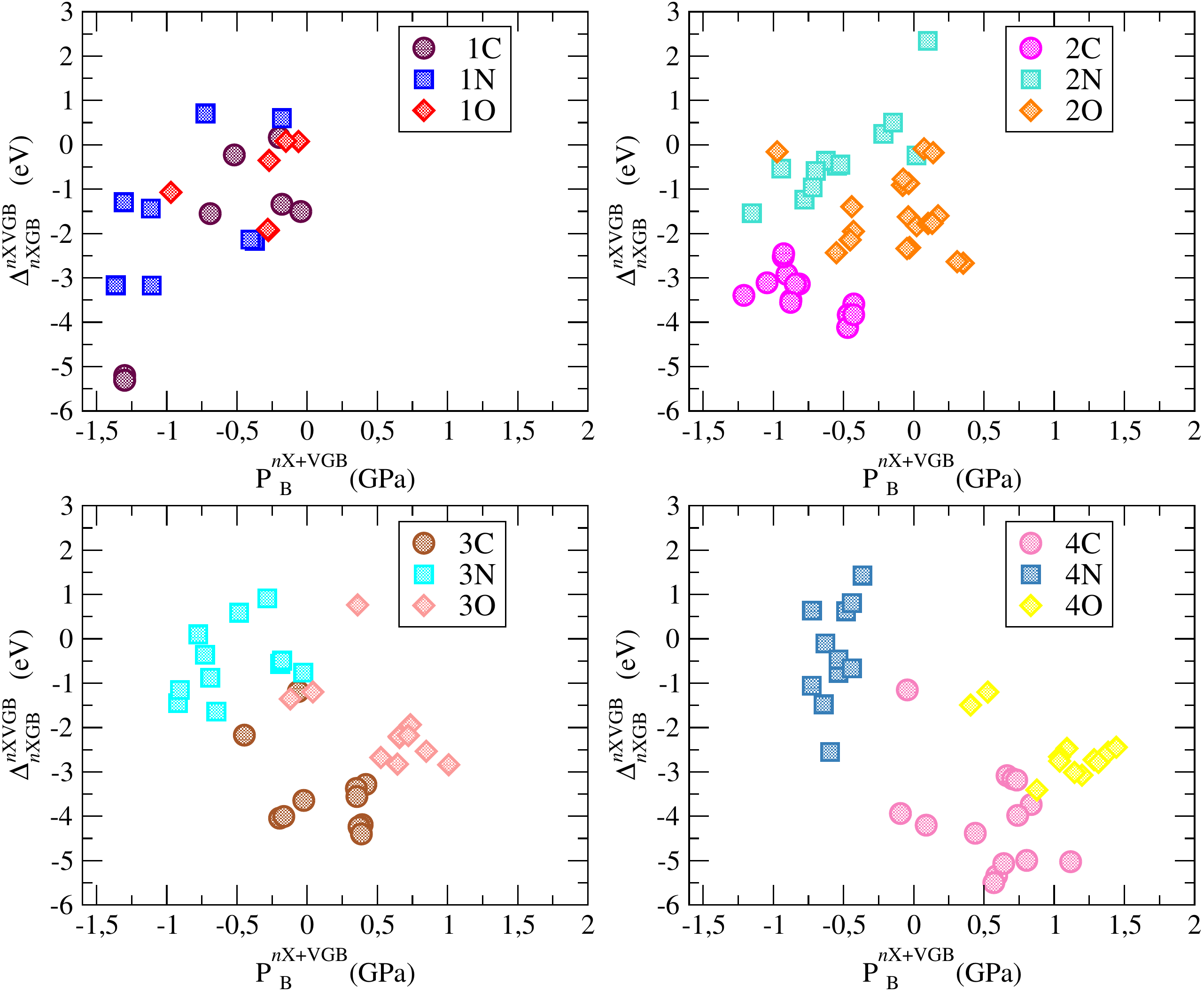}
\caption{Segregation energy $\Delta^{n\text{XVGB}}_{n\text{XGB}}$ (eV) as a function of P$^{n\text{X+VGB}}_\text{B}$ (GPa). The label X stands for the type of atoms C, N and O while $n$ indicate the number of atomic impurities. The atoms C, N and O are in the GB with a vacancy.}
\label{seg_cno_v1v2}
\end{figure}

In Fig.~(\ref{seg_cno_v1v2}) we show the segregation energies $\Delta^{n\text{XVGB}}_{n\text{XGB}}$ for the all investigated structures as a function of their pressure $P^{n\text{X+VGB}}_{\text{B}}=P_{n\text{X+VGB}} - P_{\text{B}}$ calculated as the difference between the pressure of the GB with the vacancy containing $n$ X atoms ($P_{n\text{X+VGB}}$) and the pressure of the Si bulk ($P_{\text{B}}$) . The explicit values of $\Delta^{n\text{XVGB}}_{n\text{XGB}}$ are reported in Tab.~(\ref{seg_cno_v1v2le}) for LE structures.

We observe that C is the most favourite to segregate at the GB with a vacancy. The segregation energy for 2C, 3C and 4C is always negative and the absolute value is larger than for N and O atoms. C atoms fill the voids and tend to form clusters (see Fig.~(\ref{gb_C})) without leaving dangling bonds or over-coordinated Si atoms. O has mostly a negative segregation energy. Comparing with Fig.~(\ref{seg_cno_int}), we observe that both C and O increase their capability to segregate when there is a vacancy in the GB. For Nitrogen we still have oscillations between negative and positive segregation energy. However, comparing with Fig.~(\ref{seg_cno_int}) we observe that these oscillations are present independent on the number of N atoms introduced and the absolute values are larger. It is clear that the way vacancies interact with atomic impurities can be variegated and strongly depends on the number and type of segregated species. 

\begin{figure*}[h!]
\centering
\includegraphics[scale=0.5]{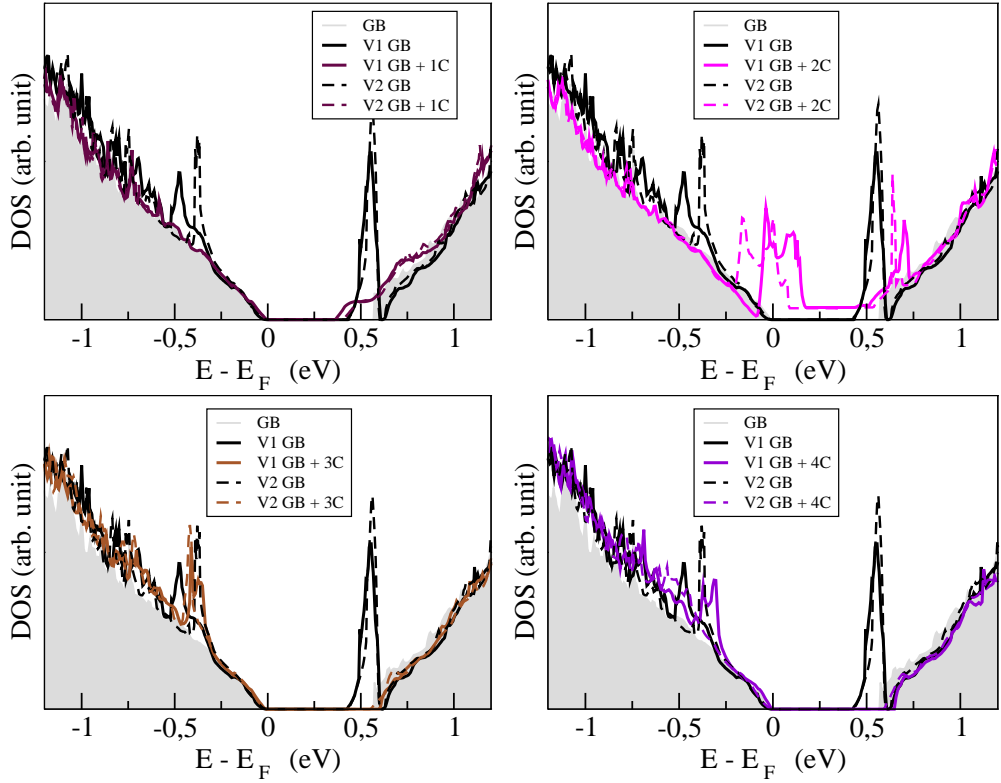}
\caption{Density of states (DOS) of $\Sigma$3\{111\} Si GB,  $\Sigma$3\{111\} Si GB with vacancies V1 and V2 and $\Sigma$3\{111\} Si GB with vacancies V1 and V2 together with 
C impurity interstitial atoms (LE). }
\label{dos_C}
\end{figure*}

To analyse the changes in the electronic properties of the GB with vacancies in the presence of C, N and O atoms we plotted the DOS in Fig.~(\ref{dos_C}). As a reference, we have also reported the DOS of only V1 and V2 GBs. These structures, without impurities, present a sharp peak at the bottom of the conduction band which, by the analysis of the projected-DOS (not shown), is due to the threefold coordinated atoms Si1 and Si2. \cite{MAJI2021116477} Another peak is also observed at around -0.5 eV in the valence band. This peak is due to four coordinated Si atoms but which undergo structural changes due to the presence of the vacancy in the GB. 


\begin{table*}[h!]
\begin{center}
\begin{tabular}{ |c |c |c |c |c |c|c|c|c| } 
\hline
X&$P^{\text{1X+VGB}}_{\text{B}}$ & $\Delta^{\text{1XVGB}}_{\text{1XGB}}$ &$P^{\text{2X+VGB}}_{\text{B}}$ & $\Delta^{\text{2XVGB}}_{\text{2XGB}}$& $P^{\text{3X+VGB}}_{\text{B}}$ &  $\Delta^{\text{3XVGB}}_{\text{3XGB}}$&$P^{\text{4X+VGB}}_{\text{B}}$ &  $\Delta^{\text{4XVGB}}_{\text{4XGB}}$ \\ 
\hline
\hline 
C  &-1.299 & -5.201 & -0.466 &-3.837 & 0.370  & -4.236  &0.590  &-5.341 \\ 
N  &-0.372 & -2.160 & -0.698 &-0.591 & -0.919  & -1.448  & -0.642  &-1.474 \\ 
O  &-0.276 & -1.933 & 0.352& -2.671 &1.008 &-2.839 & 1.198  & -3.073\\ 
\hline
\end{tabular}
\end{center}
\caption{Segregation energy $\Delta^{n\text{XVGB}}_{n\text{XGB}}$ (eV) as a function of $P^{n\text{X+VGB}}_{\text{B}}$ (GPa). The label X stands for C, N and O atoms while $n$ indicate the number of C, N and O atoms. The values are reported for the LE structures.}
\label{seg_cno_v1v2le}
\end{table*}

\begin{figure*}[h!]
\centering
\includegraphics[scale=0.5]{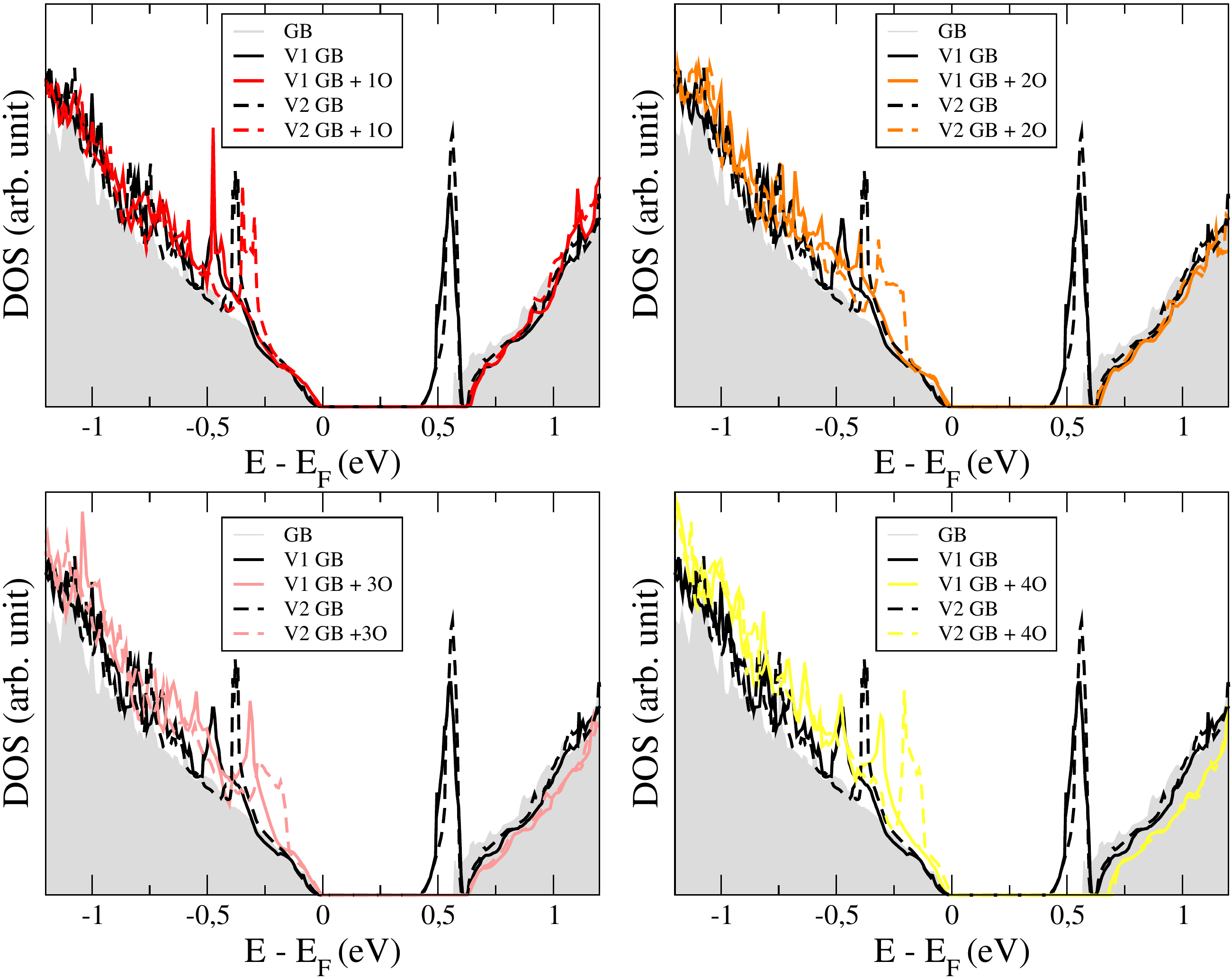}
\caption{Density of states (DOS) of $\Sigma$3\{111\} Si GB,  $\Sigma$3\{111\} Si GB with vacancies V1 and V2 and $\Sigma$3\{111\} Si GB with vacancies V1 and V2 together with 
O impurity interstitial atoms (LE). }
\label{dos_O}
\end{figure*}

The DOS for C and O atoms are respectively shown in Fig.~(\ref{dos_C}) and in Fig.~(\ref{dos_O}). The common feature of these DOS is that they restore the DOS of the pure GB. C and O atoms fill the voids by capping the dangling bonds reproducing the electronic properties of the pure GB. This effect is particularly clear increasing the number of O and C impurities. The electronic stability is supported by a stronger segregation energy. This confirms that O and C preferentially segregate at GB but  also point out that the electronic properties are not particularly modified with respect to the pure GB. This observation demonstrate that the intrinsic properties of the GB are more important than the presence of C and O impurities in order to modify the electronic properties of the devices. The case of the 2C is certainly an exception because the space around it does not allow C to form neither a simple nor a double bond and therefore the delocalised electron makes the structure gapless. However, we believe that this is very rare in experiments where usually the number of C atoms segregating is much higher.

\begin{figure*}[h!]
\centering
\includegraphics[scale=0.5]{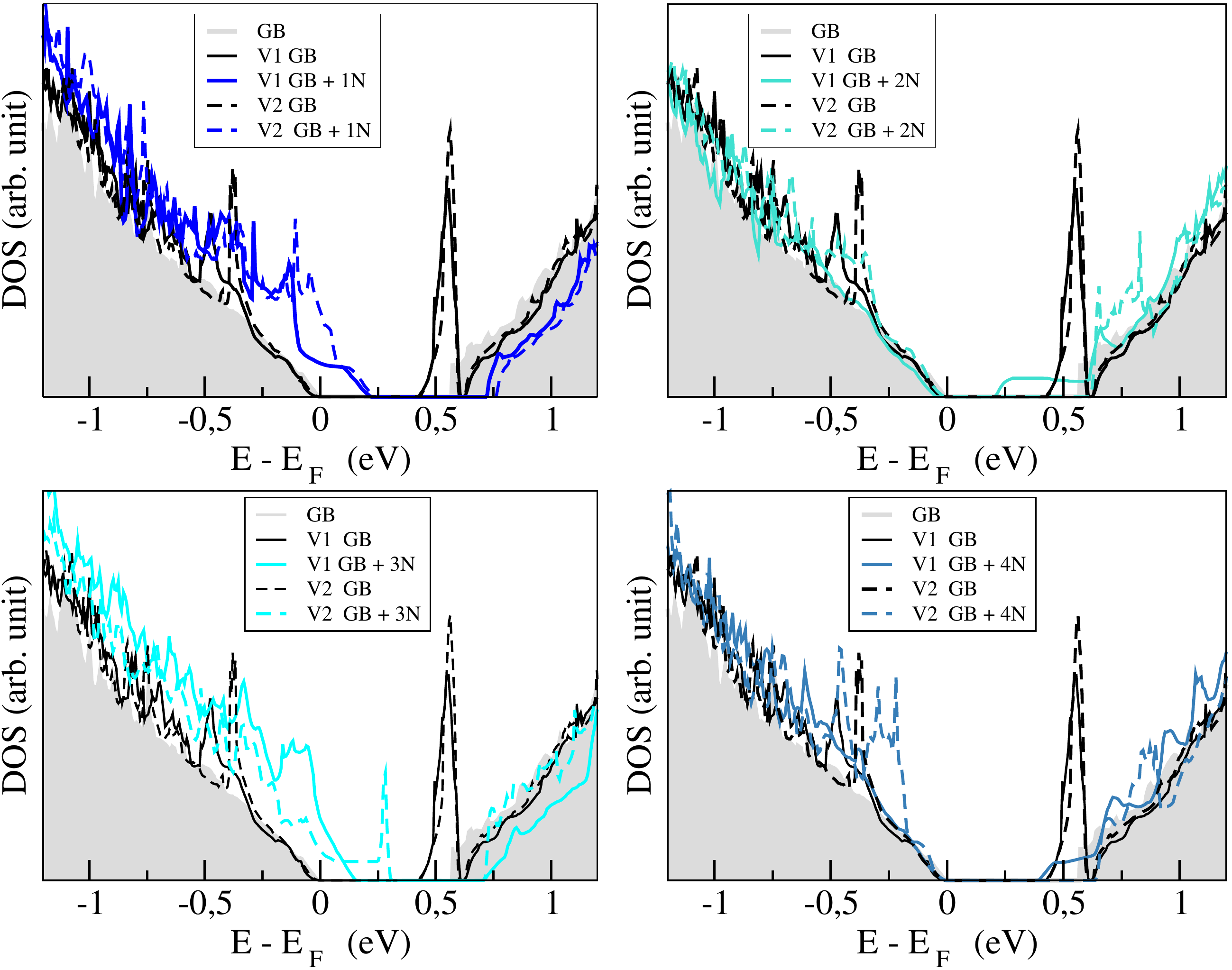}
\caption{Density of states (DOS) of $\Sigma$3\{111\} Si GB,  $\Sigma$3\{111\} Si GB with vacancies V1 and V2 and $\Sigma$3\{111\} Si GB with vacancies V1 and V2 together with 
N impurity interstitial atoms (LE). }
\label{dos_N}
\end{figure*}

The DOS of N is shown in Fig.~(\ref{dos_N}). The electronic behaviour is more difficult to understand and to predict with respect to O and C. From the analysis of the V1 structures, the 1N and 3N systems have one N that is 2-fold coordinated and therefore gapless because of the delocalisation of one electron. Instead, from the analysis of the V2 structures, the 1N has the N atom 3-fold coordinated but a Si atom that is 3-fold coordinated. The system is gapless and the states in the gap comes from that Si. In the 3N system the behaviour is the same as for V1. One N is 2-fold coordinated and the system is gapless. Both for V1 and V2, 2N and 4N atoms instead fill the voids created by the vacancies and only a flat density of electronic states from nitrogen appear at the bottom of conduction band. 

In conclusion, C, N and O segregate at GB with vacancies and the segregation is favoured increasing the number of impurities atoms introduced in the GB. However, C and O have different impacts on the electronic properties than N. In fact, C and O tend to fix the dangling bonds and the voids restoring the electronic properties of the GB without vacancies. Instead, the behaviour of N is more problematic. We obtained a completely different behaviour if we have an odd or even number of N atoms in the GB. This implies a strong difficulty in controlling the properties of the material in the presence of N impurities.

\section{Conclusions}\label{sec:conclude}

GBs are defects in the multi-crystalline Si which can induce charge carriers recombination. GBs can also easily form vacancies with deep defect electronic states and are also preferential segregation sites for various impurity species. We studied from first-principles the segregation energy of C, N and O impurity atoms at the $\Sigma$3\{111\} Si GB with and without vacancies. Moreover, we correlate the energetic properties with the structural and the electronic ones.

We obtained that when no vacancies are present, C atoms, and in particular O atoms, have a small ability to segregate. Instead, when vacancies are present both C and O strongly increase their capacity to segregate. In particular, the segregation of C becomes highly favourable. The electronic properties of the GBs in presence of O are not affected by the segregation, as it always restores a stable SiO$_2$ configuration. The same happens for C in the presence of a vacancy. In fact, C fills the voids by forming clusters without leaving dangling bonds or over-coordination of Si. However, when there is no vacancy the C atoms have not enough space to form clusters. If possible, they tend to restore a SiC structure otherwise they can be under-coordinated together with some Si atoms. These can strongly change the electronic properties with new energy states that appear in the gap. 

In the case of the segregation of N atoms, it is not possible to find a clear trend both with and without vacancies. Segregation energies oscillate between negative and positive values independently on the number of N atoms. N is not isovalent with Si as C and O. It is therefore more flexible in finding different structural arrangements within the GB. Concerning the electronic properties we observed that an odd number of N atoms gives a gapless structure, while an even number gives a semiconducting structure. 
This means that also for the electronic properties it is very difficult to find a trend. This implies a strong difficulty in controlling the properties of the material in the presence of N impurities.

The study of how the structural, energy and electronic properties are modified by GBs and defects like vacancies and impurities will allow to engineer Si solar cells with ever greater efficiency. 

 \section{SUPPLEMENTARY MATERIAL}
See the supplementary material for the projected density of
states of the studied structures.

\section{Acknowledgements}
We would like to thank the University of Modena and Reggio Emilia for the financial support (FAR dipartimentale 2020) and Centro Interdipartimentale En$\&$Tech, as well as the CINECA HPC facility for the approved ISCRA C project SiGB-NMI (IsC86\_SiGB-NMI).


\bibliography{bib}

\end{document}